\documentstyle[12pt,axodraw]{article} 
\catcode`@=11
\newcount\@tempcntc
\def\@citex[#1]#2{\if@filesw\immediate\write\@auxout{\string\citation{#2}}\fi
\@tempcnta\z@\@tempcntb\m@ne\def\@citea{}\@cite{\@for\@citeb:=#2\do
{\@ifundefined
{b@\@citeb}{\@citeo\@tempcntb\m@ne\@citea\def\@citea{,}{\bf
?}\@warning
       {Citation `\@citeb' on page \thepage \space undefined}}%
    {\setbox\z@\hbox{\global\@tempcntc0\csname b@\@citeb\endcsname\relax}%
     \ifnum\@tempcntc=\z@ \@citeo\@tempcntb\m@ne
       \@citea\def\@citea{,}\hbox{\csname b@\@citeb\endcsname}%
     \else \advance\@tempcntb\@ne \ifnum\@tempcntb=\@tempcntc
      \else\advance\@tempcntb\m@ne\@citeo
      \@tempcnta\@tempcntc\@tempcntb\@tempcntc\fi\fi}}\@citeo}{#1}}
\def\@citeo{\ifnum\@tempcnta>\@tempcntb\else\@citea\def\@citea{,}%
  \ifnum\@tempcnta=\@tempcntb\the\@tempcnta\else
   {\advance\@tempcnta\@ne\ifnum\@tempcnta=\@tempcntb \else
   \def\@citea{--}\fi
   \advance\@tempcnta\m@ne\the\@tempcnta\@citea\the\@tempcntb}\fi\fi}
   \catcode`@=12

\def\theequation{\arabic{section}.\arabic{equation}}
\begin{document}

\begin{flushright}
MPI/PhT/98--21\\ 
hep-ph/9803297\\ 
March 1998
\end{flushright}

\begin{center}
{\LARGE {\bf CP-Odd Tadpole Renormalization of }}\\[0.4cm] 
{\LARGE {\bf Higgs Scalar-Pseudoscalar Mixing }}\\[2.5cm]
{\large Apostolos Pilaftsis}\footnote[1]{E-mail address:
  pilaftsi@mpmmu.mpg.de}\\[0.4cm] 
{\em Max-Planck-Institut f\"ur
       Physik, F\"ohringer Ring 6, 80805 Munich, Germany}
\end{center}
\vskip1.7cm  \centerline{\bf  ABSTRACT}  
We  consider  an Abelian model  with   a CP-conserving Higgs potential
spanned by two  complex Higgs fields.  The  CP invariance of the Higgs
potential  is then broken  explicitly beyond the Born approximation by
introducing  soft-CP-violating  Yukawa   interactions.   Based on  the
non-renormalization  theorem, we  derive  the  consistency  conditions
under which a CP-odd counterterm exists and, at the same time, renders
the  one-loop-induced mixing of  a CP-even Higgs   boson with a CP-odd
Higgs  scalar   ultra-violet   finite.   The   novel  CP-odd   tadpole
renormalization may then    be   determined  from the     minimization
constraints   on  the   Higgs  potential.  Finally,    we discuss  the
phenomenological  consequences   of   the  so-generated   CP-violating
scalar-pseudoscalar mixing for the electric dipole moments of neutron,
electron and muon.\\[0.3cm]
PACS nos.: 11.30.Er, 13.10.+q, 14.80.Bn, 14.80.Cp

\newpage 
\setcounter{equation}{0}
\section{Introduction}

Despite  physicists'  continuous effort  since   the  discovery of  CP
violation in  the $K^0\bar{K}^0$ system  in 1964 \cite{review}, a deep
understanding of the  origin of CP asymmetry  in nature remains  still
elusive thus far.   This  fact has  rendered  the  whole  issue of  CP
nonconservation  from  the  theoretical   point   of view   even  more
challenging.  In  the  existing literature, two generically  different
scenarios have  been proposed at the Lagrangian   level to explain the
observed CP asymmetry.    In the first  scenario, complex  parameters,
such  as complex Yukawa   couplings, are introduced in the  Lagrangian
which break explicitly CP invariance.  Such a  scenario of explicit CP
violation  is realized by the  well-known Standard Model (SM) and many
of its minimal extensions.  Another  appealing scheme arises when  the
ground state of  the Higgs potential  is not  invariant under CP.   To
make such a scheme  work, one needs to extend  the Higgs sector of the
SM by adding more  than one  Higgs field.   In addition,  one requires
that the  complete tree-level Lagrangian be   CP symmetric.  After the
spontaneous  symmetry breaking   (SSB)  of the  Higgs potential,   the
resulting vacuum state  is no longer  a CP eigenstate, thereby leading
to a CP-noninvariant theory.   Such a mechanism is  called spontaneous
CP violation.  Technically, this is manifested by the fact that one of
the  vacuum  expectation values  (VEV's) of  the  Higgs fields becomes
complex \cite{Grimus}.  In this context,  it is also worth  mentioning
the  variant, in which  the spontaneous breakdown  of  the CP symmetry
occurs beyond the Born approximation  of  the Higgs potential  through
quantum mechanical  effects.  This mechanism is  known as radiative CP
violation \cite{RCPV}.

Here, we shall study another very  interesting possibility of explicit
CP violation which may naturally take place in models with an extended
Higgs sector, such as  the two-Higgs doublet  model (2HDM) and/or  the
minimal supersymmetric SM (MSSM).   For our illustrations, we consider
an  Abelian model with a  CP-conserving Higgs potential  formed by two
complex  Higgs  fields.  Such  a  model predicts  four neutral  (real)
scalars: Two physical CP-even  Higgs bosons, denoted  as $H$  and $h$,
one physical  CP-odd Higgs boson, $A$,  and the CP-odd Goldstone boson
$G^0$ which  becomes the longitudinal  component of  the massive gauge
boson $Z$.  In  this model, all the  CP-violating mixings $HA$,  $hA$,
$HG^0$ and $hG^0$ are absent at  the tree level,  and to all orders in
perturbation theory,  if CP is an exact   symmetry of the  Lagrangian. 
However,  complex Yukawa couplings of  the Higgs fields to fermions or
charged scalars  may explicitly break  CP invariance. Depending on the
detailed form of Yukawa interactions, one finds  in general that Higgs
scalar-pseudoscalar transitions induced by one-loop Feynman graphs are
{\em  not} ultra-violet  (UV)  finite.  In   a  sense this  may appear
paradoxical, since one  would  expect that the  $HA$-type  transitions
should be  UV safe by  themselves, as the tree-level CP-invariant form
of   the  Higgs   potential cannot   produce  the   necessary   CP-odd
counterterms (CT's) to cancel the UV divergences.  The latter may even
thwart the whole  renormalization  programme of  the model.   In  this
paper,   we offer  a field-theoretic solution   to the afore-mentioned
problem, for  which we  believe that  a  discussion at a  satisfactory
level is  still  missing  in  the present  literature.  By   examining
carefully   the minimization constraints  on  the  Higgs potential, we
observe  that new CP-odd tadpole  CT's do exist  which  may absorb the
above loop-induced UV  divergences.  Nevertheless, this is not  always
the case.    The non-renormalization  theorem dictates  the admissible
dimensional forms of CP-violating operators which can be introduced in
the  Lagrangian to break  CP without spoiling the renormalizability of
the model  itself.  Based on that theorem,   we derive the consistency
conditions, such that soft-CP-violating Yukawa couplings together with
a Higgs potential invariant under CP at the tree level can co-exist.

Apart from  the  theoretical interest  in providing  a self-consistent
solution  to  the problem  mentioned  above, Higgs scalar-pseudoscalar
transitions may directly be  probed at present and planned high-energy
machines \cite{PN,APRL}.  In the SM, $HZ$ mixing  is expected to occur
at  the three-loop level \cite{APRL}, as  shown in Fig.\  1, and hence
must be considered to  be not phenomenologically viable.  In contrast,
$HA$ mixing may be large within our scheme of CP  violation, as it can
be generated at the one-loop electroweak order.  Future $e^+e^-$, $pp$
and $\mu^+\mu^-$ colliders have the physics capabilities to search for
CP-violating signals  due to a non-vanishing  $HA$ mixing \cite{APRL}. 
On the other hand, the chirality  enhanced two-loop Barr-Zee mechanism
may give rise  to a large  contribution to the electric dipole moments
(EDM's) of neutron, electron and muon through  a sizeable $HA$ mixing. 
As a consequence, experimental limits on  the above EDM's place severe
bounds on the size of possible CP-violating $HA$-type operators.

The  paper is  organized  as  follows:  In  Section \ref{sec:pot}   we
describe the Lagrangian  of an  Abelian  model with two  complex Higgs
fields, in which the  Higgs potential respects   CP symmetry.  In  the
discussion, we also  include typical soft-CP-breaking operators  which
are  in agreement with   the non-renormalization theorem.   In Section
\ref{sec:tad}  we  examine the minimization   conditions  of the Higgs
potential and identify  the crucial CP-odd  tadpole CT's.  In  Section
\ref{sec:HA} we  calculate  the  one-loop induced  scalar-pseudoscalar
self-energies within a scenario of   Yukawa interactions, favoured  by
supersymmetry.  We then show how the UV divergences of the self-energy
graphs drop out  when  the CP-odd tadpole  renormalizations  are taken
into account.  Finally, in Section \ref{sec:Concl}  we conclude with a
short discussion  on the   phenomenological implications  of $HA$-type
mixings for the EDM's of neutron, electron and muon.

\setcounter{equation}{0}
\section{\label{sec:pot}Abelian two-Higgs model} 

We  will consider   a  U(1)$_Y$ model with  two   complex Higgs fields
$\Phi_1$ and $\Phi_2$, and require that  its complete Lagrangian be CP
invariant. As for the discussion of CP symmetry, many results obtained
in the Abelian model are also valid for  the respective 2HDM. Based on
the non-renormalization  theorem,  we   will   then  investigate   the
admissible forms  of soft-CP-breaking terms which  can be added in the
Lagrangian,  without  inducing  radiatively new   local operators that
violate the  CP invariance  of   the Higgs  potential and  hence   the
renormalizability of the model itself.

The  Lagrangian of the   Abelian  two-Higgs model may conveniently  be
expressed as follows:
\begin{equation}
  \label{LH}
{\cal L}_H\ =\ \sum_{i=1,2}\, (D^\mu \Phi_i )^* (D_\mu \Phi_i )\,
+\, {\cal L}_V\, ,
\end{equation}
where $D_\mu    = \partial_\mu  -  i g    A_\mu \widehat{Y}/2$  is the
covariant derivative,  with  $g$ and  $\widehat{Y}$ being   the  gauge
coupling and the  hypercharge operator of  the U(1)$_Y$, respectively. 
Furthermore, ${\cal L}_V$ is the part of the Lagrangian containing the
Higgs potential.   In  general, the fields   $\Phi_1$ and $\Phi_2$ are
responsible for endowing the observed fermions with masses through the
Higgs  mechanism.  By   the  same token  however,  they  also  lead to
potentially  large flavour-changing neutral   current couplings at the
tree level.   Glashow  and Weinberg  \cite{GW} suggested that  natural
flavour  conservation   may  be obtained   if  the  whole  Lagrangian,
including    the Yukawa sector ${\cal  L}_Y$,   is invariant under the
discrete  symmetry D: $\Phi_1  \to  \Phi_1$, $\Phi_2 \to -\Phi_2$  and
$d_R   \to   - d_R$, where     $d_R$ collectively denotes right-handed
down-type   quarks and  leptons.  Under    the  D symmetry, the  field
$\Phi_1$  couples to the up-type  family $u$, whereas $\Phi_2$ couples
to the down-type one $d$ only.

Imposing  the discrete     symmetry D  on   the  general   form of   a
U(1)$_Y$-invariant Higgs potential yields
\begin{eqnarray}
 \label{LV}
{\cal L}_V &=& \mu^2_1 (\Phi_1^*\Phi_1)\, +\, \mu^2_2 (\Phi_2^*\Phi_2)\, 
  +\, \lambda_1 (\Phi_1^*\Phi_1)^2\, +\, \lambda_2 (\Phi_2^*\Phi_2)^2
                                                             \nonumber\\ 
&& +\, \lambda_{34} (\Phi_1^*\Phi_1 \Phi_2^*\Phi_2)\, +\, 
\lambda_5 (\Phi_1^*\Phi_2)^2\, +\, \lambda^*_5 (\Phi_2^*\Phi_1)^2\, ,
\end{eqnarray}
with the hypercharge assignment  $Y (\Phi_1) = Y (\Phi_2)  = 1$.  Note
that the   term  proportional  to $\lambda_{34}$ also    comprises the
D-symmetric  combination    $(\Phi_1^*\Phi_2)  (\Phi_2^*\Phi_1)$.   In
addition, we readily see that ${\cal L}_V$ remains invariant under the
generalized CP transformations compatible with D symmetry
\begin{equation}
  \label{CPtrafo}
\Phi_1 \to e^{i\phi_1} \Phi_1^*\, , \qquad 
\Phi_2 \to e^{i\phi_2} \Phi_2^*\, ,
\end{equation}
provided the phases  $\phi_1$  and $\phi_2$  are chosen in  a way such
that $\phi_1 - \phi_2 =  \mbox{arg} \lambda_5$.  As a consequence, the
potential ${\cal L}_V$ is CP invariant.

We should now notice that the D symmetry of the Higgs potential cannot
be promoted to  a continuous one  of the  Peccei-Quinn type \cite{PQ},
unless $\lambda_5  = 0$.  To  be specific,  for vanishing $\lambda_5$,
one may choose the $Q$ quantum numbers for the fields
\begin{eqnarray}
  \label{Qtrafo}
Q (\Phi_1 ) &=& 2\, ,\qquad   Q (\Phi_2 )\ =\ 1\, ,\qquad
Q (u_L)\ =\ Q (d_L)\ =\ 2\, ,\nonumber\\
Q (u_R) &=& 0\, ,\qquad Q (d_R)\ =\ 1\, ,  
\end{eqnarray}
and  then show that in  addition to the   gauge symmetry U(1)$_Y$, the
total  Lagrangian ${\cal L}  = {\cal L}_H   + {\cal L}_Y$ is invariant
under U(1)$_Q$, with
\begin{equation}
  \label{LY}
-\, {\cal L}_Y \ =\ \Phi_1 \bar{u}_L h_u u_R\, +\, 
\Phi_2 \bar{d}_L h_d d_R\ +\ \mbox{H.c.}
\end{equation}
D symmetry is obtained for the  choice of the global (phase) parameter
$\phi  = \pi$.   In Eq.\  (\ref{LY}), $h_u$  and  $h_d$ are in general
(dimensionless) complex Yukawa matrices  for the up-type and down-type
families, respectively.   However, using the  freedom of re-definition
of the fermionic fields, one  can make both  matrices $h_u$ and  $h_d$
diagonal  with  positive  entries,  without   spoiling U(1)$_Q$ and  D
symmetries.  In fact, after SSB, the Yukawa  couplings $h_u$ and $h_d$
get directly  related to the observed  fermion masses.  Obviously, the
Lagrangian   ${\cal L}$   is  exactly  CP  invariant,  so  absence  of
CP-violating mixing terms of the kind $\Re e(\Phi_1) \Im m(\Phi_1)$ is
guaranteed to  all  orders in perturbation  theory.    Also, the above
local  operator being proportional to $\Im  m (\Phi_1)^2$ is forbidden
because  of  U(1)$_Y$  invariance.  We reach   the same conclusion  if
$\lambda_5  \not  = 0$.   Notwithstanding   the fact that U(1)$_Q$  is
broken    hard  by  $(\Phi_1^*\Phi_2)^2$,   it is    however  the only
four-dimensional operator that  can be generated  at high orders which
is simultaneously  invariant under D  and U(1)$_Y$.  As it should, the
effective potential is CP  invariant.   In general, this  result  will
hold true  even   after SSB.\footnote[1]{Exception  to   this case are
  scenarios of spontaneous or radiative  CP violation.  Here, we shall
  not consider such alternatives.    More   details may be found    in
  \cite{RCPV}.}

We now  examine the consequences for  the effective Higgs potential if
CP-violating Yukawa interactions are added to the Lagrangian which are
invariant under D or  U(1)$_Q$ symmetries.  Again, U(1)$_Y$ invariance
forbids the presence of  the CP-violating operator $\Re  e(\Phi_1) \Im
m(\Phi_1)$. However, after the SSB of the gauge and global symmetries,
CP-violating $HA$-type transitions may become  possible.  On the other
hand, the non-renormalization theorem  \cite{Nonren} dictates the form
of CT's for a spontaneously broken theory.   According to the theorem,
the CT's are  entirely determined  from  those given in  the symmetric
phase of the theory.  As a  consequence, the loop-induced $HA$ mixings
must be UV finite. A  scenario of this type has  been discussed in the
recent literature \cite{APRL}. Specifically, in  addition to the Higgs
scalars,  the model contains  two   iso-singlet neutrinos, denoted  as
$N_1$ and $N_2$, and   one  sequential left-handed neutrino $\nu_L$.   
Thus, the Yukawa  sector  of  the model  allows for  the  simultaneous
presence of Dirac and Majorana mass terms, {\em i.e.},
\begin{equation}
  \label{LMaj}
-{\cal L}^M_Y\ =\ \Phi_1 \bar{\nu}_L (h_1 N_1\, +\, h_2 N_2)\ +\ 
M_1 N^T_1 C N_1\ +\ M_2 N^T_2 C N_2\ +\ \mbox{H.c.}, 
\end{equation}
where $C$ is  the operator of  charge conjugation.   Lagrangian ${\cal
  L}^M_Y$ has been written in the weak basis, in which $M_1$ and $M_2$
are real,  while  $h_1$ and   $h_2$ are complex  numbers.   The  model
predicts three heavy Majorana neutrinos, which may  have masses as low
as hundreds of GeV;  they may be discovered  in the next generation of
colliders.   For the  general  case of  non-degenerate  heavy Majorana
neutrinos, the non-vanishing of  the rephasing-invariant quantity $\Im
m (h_1 h^*_2)$ signifies CP  violation.  Notice that ${\cal L}^M_Y$ is
invariant under  the  symmetries D  and U(1)$_Q$.  Fig.\   2 shows the
flavour diagrams  responsible for generating the CP-violating operator
$\Re e(\Phi_1) \Im m(\Phi_1)$ after SSB. As can  also be seen by doing
a    naive power  counting  in    Fig.\   2, the  resulting  $HA$-type
self-energies are UV finite.  For more  details the reader is referred
to \cite{APRL}.  Finally, we should remark that the $HZ$ mixing in the
minimal SM is very  analogous to the  afore-mentioned case of explicit
CP violation (see Fig.\ 1).

In the following, we will focus our attention on scenarios of explicit
CP violation, in   which the symmetries D  and/or  U(1)$_Q$ are softly
broken.  Our  main interest is  to find the consistency  conditions of
renormalization which ensure  the co-existence  of a {\em  tree-level}
CP-invariant  Higgs potential  together with  soft-CP-breaking  Yukawa
interactions.  The first soft-breaking term one may think of adding to
the potential ${\cal    L}_V$ in Eq.\  (\ref{LV})  is  $\mu^2 \Phi^*_1
\Phi_2$, where $\mu$  is in general  a complex parameter. However, the
inclusion    of  such  a two-dimensional  operator   would   lead to a
CP-violating Higgs potential already at the tree-level, unless
\begin{equation}
  \label{CPcond}
\Im m (\mu^4 \lambda_5^* ) \ =\ 0\, .
\end{equation}
Besides the  option  of fine-tuning,  the most natural  way to fulfill
CP-invariance condition  (\ref{CPcond})  is to assume  that  the Higgs
potential has  originally a global U(1)$_Q$  symmetry  which is softly
broken afterwards  by the above $\mu^2$-dependent  mass term.  In this
case, the quartic coupling  $\lambda_5$ is absent  from the potential. 
There is also  another  reason advocating for  the  naturality  of the
latter  scenario.  If we  had    broken U(1)$_Q$ by trilinear   Yukawa
operators of dimension  three,  we would  have  been compelled,  as  a
consequence  of   the non-renormalization    theorem, to  include  all
possible    operators  of   lower     dimensions,  {\em i.e.},     the
two-dimensional mass  term $\mu^2$.  Again,  for $\lambda_5  \not= 0$,
this would have   led to a  CP-violating  Higgs potential at  the tree
level.  For  the Abelian  two-Higgs  model, we can therefore  conclude
that in order to have a  {\em consistent} CP-invariant Higgs potential
at the  tree level, it  is sufficient to  require for all operators of
dimension four to be U(1)$_Q$ symmetric in the complete Lagrangian and
allow only for soft-breaking of  U(1)$_Q$ by mass and trilinear Yukawa
terms.

Motivated by  the MSSM, we now  present a  specific scenario, in which
both symmetries U(1)$_Q$ and CP are softly broken on the same footing.
The model includes  the charged scalars $\chi^\pm_L$ and $\chi^\pm_R$,
such   as  left-handed  and   right-handed scalar  quarks,  which have
trilinear Yukawa couplings  to the Higgs  fields.  To be  precise, the
interactions of the charged scalars are governed by the Lagrangian
\begin{equation}
  \label{Lchi}
-{\cal L}_Y^\chi\ =\ m^2_L \chi^+_L\chi^-_L\ +\ 
m^2_R \chi^+_R\chi^-_R\ +\ (f \Phi_1\, +\, h\Phi_2 )\chi^+_L\chi^-_R\
+\ \mbox{H.c.}\, ,
\end{equation}
where $f$ and $h$ are in general  complex mass parameters, while $m_L$
and $m_R$ are real.  We assign to the newly introduced charged scalars
the following  hypercharge  and $Q$  quantum numbers:  $Y(\chi^-_L)  =
Y(\chi^+_R) = 1/2$, and $Q(\chi^-_L) = 2$ and $Q(\chi^\pm_R) = 0$.  It
is then  clear that both CP  and U(1)$_Q$ symmetries are softly broken
by the operator $\Phi_2\chi^+_L\chi^-_R$ of dimension three.  In fact,
the  model   admits   CP violation,    unless  the rephasing-invariant
constraint $\Im m (\mu^2 f h^*) = 0$ holds true.

As we  will see in Section \ref{sec:HA},  the above scalar model leads
to Higgs scalar-pseudoscalar    self-energies which are  {\em not}  UV
finite. In the next  section, we shall show  how a CP-invariant  Higgs
potential  can  still  produce the  necessary CP-odd  CT's,  which can
cancel the loop-induced UV divergences.

\setcounter{equation}{0}
\section{\label{sec:tad}CP-odd tadpole renormalization} 

In this section, we shall  examine the minimization conditions imposed
on a Higgs potential  which is invariant under CP  at the tree level.  
We  shall  then show  that CP-odd   tadpole  CT's do  exist which  are
relevant for the renormalization of Higgs scalar-pseudoscalar mixings.
Following  the discussion given  in Section \ref{sec:pot}, we consider
the Higgs potential
\begin{eqnarray}
  \label{Lmu}
{\cal L}^\mu_V &=& \mu^2_1 (\Phi_1^*\Phi_1)\, +\, \mu^2_2 (\Phi_2^*\Phi_2)\, 
+\, \mu^2 (\Phi_1^*\Phi_2)\, +\, \mu^{*2} (\Phi_2^*\Phi_1) \nonumber\\ 
&&+\, \lambda_1 (\Phi_1^*\Phi_1)^2\, +\, \lambda_2
(\Phi_2^*\Phi_2)^2\, +\, \lambda_{34} (\Phi_1^*\Phi_1
\Phi_2^*\Phi_2)\, +\, \dots
\end{eqnarray}
Unlike the mass term $\mu^2$, the  Higgs potential ${\cal L}^\mu_V$ is
symmetric  under U(1)$_Q$; $\mu^2$  breaks only  softly U(1)$_Q$.  The
ellipses in Eq.\ (\ref{Lmu})  denote possible new quartic interactions
of the type $\Phi^*_1\Phi_1 \chi^+_L\chi^-_L$, $(\chi^+_L\chi^-_L)^2$,
{\em etc.}, which   are allowed by U(1)$_Q$  and  U(1)$_Y$.  Since the
charged scalars $\chi^\pm_L$   and $\chi^\pm_R$ do not develop  VEV's,
the presence of these additional quartic operators will not affect the
minimization constraints on the Higgs potential.  For phenomenological
simplicity, we may   assume  that these extra quartic   couplings  are
rather suppressed.  Finally, it is very interesting to notice that the
Higgs potential in Eq.\ (\ref{Lmu}) is analogous  to that predicted in
the  MSSM \cite{HiggsGuide}.   So, the  discussion  presented in  this
section can easily carry over to the latter case as well.

Without any  loss of generality, it proves  more convenient to perform
the minimization  of the Higgs potential  in the weak  basis, in which
both VEV's of the  Higgs  fields are  positive.  As usual, we use  the
linear parameterizations
\begin{equation}
  \label{Phis}
\Phi_1\ =\ \frac{1}{\sqrt{2}}\, (v_1\, +\, H_1\, +\, iA_1)\, ,\qquad 
\Phi_2\ =\ \frac{1}{\sqrt{2}}\, (v_2\, +\, H_2\, +\, iA_2)\, ,
\end{equation}
where  $v_1$ and  $v_2$  are the VEV's  of  the unbroken Higgs fields,
$H_1$ and  $H_2$ are CP-even  Higgs  bosons, and $A_1$  and $A_2$  are
CP-odd scalars.  The minimization  constraints may then  be determined
by demanding the vanishing of the following tadpole parameters:
\begin{eqnarray}
  \label{TH1}
T_{H_1} &\equiv& \frac{\partial {\cal L}^\mu_V}{\partial H_1}
\bigg|_{\langle H_i\rangle =\langle A_i\rangle =0}\ =\ 
v_1\, \Big( \mu^2_1\, +\, \Re e \mu^2\, \frac{v_2}{v_1}\, +\,
\lambda_1 v^2_1\, +\, \frac{1}{2}\, \lambda_{34}v^2_2 \Big)\, ,\\
  \label{TH2}
T_{H_2} &\equiv& \frac{\partial {\cal L}^\mu_V}{\partial H_2}
\bigg|_{\langle H_i\rangle =\langle A_i\rangle =0}\ =\ 
v_2\, \Big( \mu^2_2\, +\, \Re e \mu^2\, \frac{v_1}{v_2}\, +\,
\lambda_2 v^2_2\, +\, \frac{1}{2}\, \lambda_{34}v^2_1 \Big)\, ,\\
  \label{TA1}
T_{A_1} &\equiv& \frac{\partial {\cal L}^\mu_V}{\partial A_1}
\bigg|_{\langle H_i\rangle =\langle A_i\rangle =0}\ =\ v_2 \Im
m\mu^2\, ,\\
  \label{TA2}
T_{A_2} &\equiv& \frac{\partial {\cal L}^\mu_V}{\partial A_2}
\bigg|_{\langle H_i\rangle =\langle A_i\rangle =0}\ =\ -v_1 \Im
m\mu^2\, .
\end{eqnarray}
Clearly, in  the Born approximation one has  that $\Im m\mu^2 = 0$ and
CP is a good symmetry of the Higgs  potential. However, at high orders
$\Im m\mu^2$ does    no   longer vanish due  to   CP-violating  Yukawa
interactions. In  fact,  the CP-odd  tadpoles $T_{A_1}$  and $T_{A_2}$
depend explicitly  on $\Im m\mu^2$ and are   therefore very crucial to
render the $H_iA_j$ self-energies UV finite.

It is now  important to identify the true  Goldstone boson,  $G^0$, of
the theory, which is associated with the SSB of U(1)$_Y$. According to
the Goldstone  theorem, $G^0$ should remain  massless  and have a pure
pseudoscalar coupling   to like-flavour fermions   to  all  orders  in
perturbation theory. If the Abelian U(1)$_Y$ symmetry is gauged, $G^0$
becomes the  longitudinal degree of freedom of  the gauge boson  $Z$. 
Apart from   obtaining  the massless eigenstate   from the  Higgs mass
matrix, the easiest  way  to  find $G^0$  is   to look for  the   flat
direction of the  Higgs potential. This  amounts to finding  the field
configuration $G^0$ for which
\begin{equation}
  \label{G0}
\frac{\partial {\cal L}^\mu_V}{\partial G^0}
\bigg|_{\langle H_i\rangle =\langle A_i\rangle =0}\ \equiv\ T_{G^0}\
=\  0\, .
\end{equation}
Since $G^0$ is CP-odd,  it must be a  linear combination of the CP-odd
fields $A_1$ and $A_2$.  Thus, we are seeking  for  a solution to  the
equation
\begin{eqnarray}
  \label{TG0}
T_{G^0} &=& \frac{\partial A_1}{\partial G^0}\, T_{A_1}\ +\ 
                \frac{\partial A_2}{\partial G^0}\, T_{A_2}\nonumber\\
&=& c_\beta T_{A_1}\ +\ s_\beta T_{A_2}\ =\ 0\, ,
\end{eqnarray}
where the usual short-hand notations $s_x = \sin x$ and $c_x = \cos x$
are employed. Eq.\ (\ref{TG0}) implies that
\begin{equation}
  \label{tanbeta}
\tan\beta\ =\ -\, \frac{T_{A_1}}{T_{A_2}}\ =\ \frac{v_2}{v_1}\, .    
\end{equation}
As a result, the two physical CP-odd states $G^0$ and $A$ are related
to $A_1$ and $A_2$ through the orthogonal transformation
\begin{equation}
  \label{G0A}
\left( \begin{array}{c} A_1 \\ A_2\end{array}\right)\ =\
\left( 
\begin{array}{cc} c_\beta & -s_\beta \\ s_\beta & c_\beta\end{array}\right)\,
\left( \begin{array}{c} G^0 \\ A\end{array}\right)\, .
\end{equation}
The very same  result would have  been obtained if we had diagonalized
the mass matrix  for   the pseudoscalar bosons.   Likewise,  the Higgs
scalars  $H_1$ and $H_2$  are  related to  the physical CP-even  Higgs
particles,  denoted as $h$  and  $H$, through an analogous  orthogonal
rotation of angle $\theta$, {\em i.e.},
\begin{equation}
  \label{H1H2}
\left( \begin{array}{c} H_1 \\ H_2\end{array}\right)\ =\
\left( 
\begin{array}{cc} c_\theta & -s_\theta \\ s_\theta & c_\theta
\end{array}\right)\,
\left( \begin{array}{c} h \\ H\end{array}\right)\, .
\end{equation}
Details on  the   diagonalization of  the   Higgs mass   matrices  and
discussion of stability conditions for  the Higgs potential are  given
in the Appendix. 

Because of the  above orthogonal transformation  of the CP-odd fields,
it  is  obvious that the number   of independent tadpole parameters in
Eqs.\ (\ref{TH1})--(\ref{TA2})  has  been reduced to  three.  However,
there still exists one non-trivial CP-odd tadpole CT, given by
\begin{eqnarray}
  \label{TA}
T_A &=& \frac{\partial A_1}{\partial A}\, T_{A_1}\ +\ 
                \frac{\partial A_2}{\partial A}\, T_{A_2}\nonumber\\
&=& -s_\beta T_{A_1}\ +\ c_\beta T_{A_2}\ =\ -\, v\, \Im m \mu^2\, ,
\end{eqnarray}
with $v = \sqrt{v^2_1 + v^2_2}$. In particular, we find that all Higgs
scalar-pseudoscalar mixings in the Higgs potential are proportional to
the tadpole renormalization constant  $T_A$  of the pseudoscalar  $A$. 
More explicitly, we obtain the $HA$-type CT's
\begin{equation}
  \label{LHA}
{\cal L}_V^{HA} \ =\  \frac{T_A}{v}\, \Big[\, 
(c_\theta h - s_\theta H)(s_\beta G^0 + c_\beta A)\, -\,
(s_\theta h + c_\theta H)(c_\beta G^0 - s_\beta A)\,\Big]\, .
\end{equation}
{}From the above discussion, it became clear that CP-violating quantum
effects will shift the Higgs ground states to a CP-odd direction which
should be re-adjusted by requiring that the CT $T_A$ should cancel the
loop-induced tadpole graph  of  the $A$  boson. In this  way,  a novel
CP-odd  renormalization is obtained    which must be  included in  the
calculation of $HA$-type self-energies.  In  the next section, we will
illuminate this point further.

\setcounter{equation}{0}
\section{\label{sec:HA}Loop-induced scalar-pseudoscalar mixing} 

To elucidate the  necessity  of a CP-odd tadpole  renormalization,  we
shall calculate $HA$-type  self-energies  within an extension  of  the
two-Higgs  model, in   which   the charged  scalars  $\chi^\pm_L$  and
$\chi^\pm_R$ are  introduced.    As  has  been  discussed  in  Section
\ref{sec:pot},  such   a  scenario  can  consistently   accommodate  a
CP-invariant Higgs potential  at  the   tree  level together with    a
CP-violating  Yukawa  sector  described   by  the   Lagrangian  ${\cal
  L}^\chi_Y$ in Eq.\ (\ref{Lchi}).

{}From  Eq.\ (\ref{Lchi}), we may  now determine  the mass eigenstates
$\chi^\pm_1$ and $\chi^\pm_2$ for the charged scalars, {\em i.e.},
\begin{equation}
  \label{Mchi}
-{\cal L}^\chi_{\rm mass}\ =\ (\chi^+_L,\ \chi^+_R)\, 
\left( \begin{array}{cc} m^2_L & a^2 \\ a^{*2} & m^2_R \end{array}
\right)\, \left( \begin{array}{c} \chi^-_L \\ \chi^-_R
\end{array}\right)\ = \
(\chi^+_1,\ \chi^+_2)\, 
\left( \begin{array}{cc} M^2_1 & 0 \\ 0 & M^2_2 \end{array}
\right)\, \left( \begin{array}{c} \chi^-_1 \\ \chi^-_2
\end{array}\right)\, , 
\end{equation}
with $a^2 = (f v_1 + hv_2)/\sqrt{2}$ and 
\begin{equation}
  \label{chi1chi2}
\left( \begin{array}{c} \chi^-_L \\ \chi^-_R\end{array}\right)\ =\
\left( 
\begin{array}{cc} c_\phi & s_\phi e^{i\delta} \\ 
-s_\phi e^{-i\delta} & c_\phi \end{array}\right)\,
\left( \begin{array}{c} \chi^-_1 \\ \chi^-_2 \end{array}\right)\, .
\end{equation}
The requirement of having positive masses for charged scalars leads to
the  inequality $m_Lm_R -  |a|^2 > 0$.   In Eq.\ (\ref{chi1chi2}), the
phase $\delta$ is  trivial, since it can always  be  eliminated by the
judicious phase  re-definition  of the  field $\chi^-_R$, {\em  e.g.},
$\chi^-_R  \to e^{-i\delta} \chi^-_R$. In  this weak basis, $a^2$ is a
real parameter.   The other mass  parameters in  Eq.\ (\ref{Mchi}) are
related to the physical  masses   of the  charged scalars,  $M_1$  and
$M_2$, and the mixing angle $\phi$ as follows:
\begin{eqnarray}
  \label{Massrel}
m^2_L & =& M^2_1 c^2_\phi\, +\, M^2_2 s^2_\phi\, ,\nonumber\\
m^2_R & =& M^2_1 s^2_\phi\, +\, M^2_2 c^2_\phi\, ,\nonumber\\
a^2 & = & (M^2_2 - M^2_1)\, s_\phi c_\phi\, .
\end{eqnarray}

As has been mentioned in  Section \ref{sec:pot}, the model violates CP
through trilinear Yukawa interactions in  ${\cal L}^\chi_Y$ for $\Im m
(\mu^2 fh^*) \not= 0$. In particular, one finds that the CP-even Higgs
bosons $H_1$  and $H_2$ (or equivalently  $h$  and $H$) couple  to the
same bilinear operators of charged scalars as  the CP-odd bosons $G^0$
and   $A$  do.  These   couplings  are  precisely  those  which induce
$HA$-type  transitions at the one-loop   level.   To be specific,  the
interaction Lagrangian of interest,  obtained from ${\cal  L}^\chi_Y$,
reads:
\begin{eqnarray}
  \label{Lint}
{\cal L}^\chi_{\rm int} &=& -\,\frac{i}{v}\, a^2\, G^0\chi^+_1\chi^-_2
      \ +\ \frac{2s_\phi c_\phi}{vs_\beta c_\beta}\, 
      \Im m b^2\, A (\chi^+_1\chi^-_1 - \chi^+_2\chi^-_2)\nonumber\\
&&-\, \frac{i}{vs_\beta c_\beta}\, ( a^2 c^2_\beta\, -\, 
  b^2 c^2_\phi\, -\, b^{*2} s^2_\phi )\, 
  A \chi^+_1\chi^-_2\ +\ \frac{2s_\phi c_\phi}{vc_\beta}\,
\Re e b^2\, H_1 (\chi^+_1\chi^-_1 - \chi^+_2\chi^-_2)\nonumber\\
&&+\, \frac{2s_\phi c_\phi}{vs_\beta}\,
( a^2 - \Re eb^2 )\, H_2 (\chi^+_1\chi^-_1 - \chi^+_2\chi^-_2)\
-\ \frac{1}{v c_\beta}\, ( b^2 c^2_\phi - b^{*2} s^2_\phi )\,
H_1 \chi^+_1\chi^-_2\nonumber\\
&&-\, \frac{1}{vs_\beta}\, \Big[ a^2 (c^2_\phi - s^2_\phi) - b^2 
c^2_\phi + b^{*2}s^2_\phi\, \Big]\, H_2 \chi^+_1\chi^-_2\ +\ \mbox{H.c.},
\end{eqnarray}
where the   mass parameter $b^2  =  f  v_1/\sqrt{2}$ may  take complex
values.  To avoid double counting  in Eq.\ (\ref{Lint}), the Hermitian
conjugate (H.c.)  term is understood to be included only for couplings
which are  not    self-conjugate by themselves.  Moreover,    the weak
eigenstates $H_1$ and $H_2$ may be expressed  in terms of the physical
states  $h$ and $H$ by virtue  of Eq.\ (\ref{H1H2}).  Nevertheless, we
can   simplify  our  calculations   by   assuming  that  the   unknown
Higgs-scalar mixing $\theta$ is very  small.  In the limit of  $\theta
\to 0$, we then have $h \equiv H_1$ and $H\equiv H_2$.

Since our interest is to study the UV behaviour of scalar-pseudoscalar
transitions, we calculate  the loop-induced $HA$-type self-energies at
vanishing  external momentum, {\em i.e.},  $q\to 0$.  This may also be
justified by  the effective  potential formalism \cite{EPF},  in which
the charged  scalars  are integrated out  as  being heavy   degrees of
freedom.  As an example, let  us consider the one-particle-irreducible
(1PI) self-energy graph shown in  Fig.\ 3(a), which  gives rise to the
$G^0h$ mixing  at the one-loop  level.  It is then straightforward  to
obtain
\begin{equation}
  \label{PiG0H1}
\Pi^{G^0h}_{(a)} (q^2=0)\ =\ -2\, \frac{s_\phi c_\phi}{c_\beta}\, 
\frac{\Im mb^2}{v^2}\, (M^2_1 - M^2_2)\, I(M_1,M_2)\, ,
\end{equation}
where the loop function,
\begin{eqnarray}
  \label{IM1M2}
I(M_1,M_2) &=& \mu^{4-n}\int\!\! \frac{d^n k}{(2\pi)^n i}\, 
         \frac{1}{(k^2 - M^2_1)(k^2 -M^2_2)} \nonumber\\
&=& \frac{1}{16\pi^2}\, \Big[\, \frac{1}{\varepsilon}\, -\, 
\gamma_E\, +\, 1\, +\, \ln 4\pi\, -\, \ln\Big( \frac{M_1
  M_2}{\mu^2}\Big) \nonumber\\
&& +\, \frac{M^2_1 + M^2_2}{2(M^2_1 - M^2_2)}\, 
\ln\Big(\frac{M^2_2}{M^2_1}\Big)\, \Big]\, ,
\end{eqnarray}
is defined in $n = 4 - 2\varepsilon$  dimensions.  The parameter $\mu$
in Eq.\ (\ref{IM1M2}) denotes the 't Hooft mass.   As a consequence of
the   Goldstone theorem mentioned above,  the  $G^0h$ self-energy must
vanish for zero momentum transfer.  Otherwise, one would find that the
true  Goldstone boson $G^0$    receives a non-zero radiative   mass in
contradiction with the   Goldstone theorem, since $\Pi^{G^0G^0} (0)  =
[\Pi^{G^0h}_{(a)}(0)]^2/M^2_h  \not= 0$  through $G^0h$ mixing  at two
loops.  In fact,  for non-degenerate charged scalars,  the self-energy
$\Pi^{G^0h}_{(a)}$  does  not vanish; it   is  even plagued  by  an UV
infinity.

Evidently,  one is  faced  with the fact that   some CP-odd CT must be
included in the calculation, as  shown in Fig.\  3(b), so as to render
$G^0h$   self-energy UV    finite.  Fortunately,   Lagrangian   ${\cal
  L}^{HA}_Y$      in   Eq.\ (\ref{LHA})     provides   the   necessary
renormalization  constant  for    that  purpose.  Indeed,    for  zero
Higgs-scalar mixing ($\theta =0$), we have
\begin{equation}
  \label{PiTA}
\Pi^{G^0h}_{(b)} (0)\ =\ \frac{s_\beta T_A}{v}\ .
\end{equation}
The  CP-odd tadpole  parameter  $T_A$ may  be determined  by the usual
renormalization condition
\begin{equation}
  \label{TAcond}
\Gamma^A (0)\ +\ T_A\ =\ 0\, ,
\end{equation}
{\em  i.e.}, quantum corrections must  not shift the true ground state
of the effective potential.  In Eq.\ (\ref{TAcond}), $\Gamma^A (0)$ is
the  1PI tadpole graph  of the CP-odd  Higgs scalar $A$, which is also
depicted in Fig.\ 3(c). In this way, we find
\begin{equation}
  \label{TAform}
T_A\ =\ 2\, \frac{s_\phi c_\phi}{s_\beta c_\beta}\, 
\frac{\Im mb^2}{v}\, (M^2_1 - M^2_2)\, I(M_1,M_2)\, .
\end{equation}
It is then not difficult to see that 
\begin{equation}
  \label{G0hzero}
\Pi^{G^0h}_{(a)} (0)\ +\ \Pi^{G^0h}_{(b)} (0)\ =\ 0\, ,
\end{equation}
as it should on account of the Goldstone theorem.

By analogy,  we  can calculate   the  $hA$ self-energy.  The  diagrams
contributing  to  such   a CP-violating  Higgs-scalar  transition  are
displayed in Fig.\ 4.   The crucial difference  with the $G^0h$ mixing
in  Fig.\  3   is  that  in   addition to the   off-diagonal couplings
$h\chi^\pm_1\chi^\mp_2$  and  $A\chi^\pm_1 \chi^\mp_2$, the transition
$hA$ can  also  proceed  via the   diagonal  couplings $h\chi^+_{1(2)}
\chi^-_{1(2)}$   and   $A\chi^+_{1(2)} \chi^-_{1(2)}$.  The individual
contributions shown in Fig.\ 4(a)--(c) are given by
\begin{eqnarray}
  \label{hAa}
\Pi^{hA}_{(a)} (0) & = & -2\, \Big[\, \frac{s_\phi c_\phi}{s_\beta}\, 
    \frac{\Im m b^2}{v^2}\, (M^2_1 - M^2_2)\ +\
    4\, \frac{s^2_\phi c^2_\phi}{s_\beta c^2_\beta}\,
    \frac{\Im m b^2\Re e b^2}{v^2}\, \Big]\, I(M_1,M_2)\, ,\\
  \label{hAb}
\Pi^{hA}_{(b)} (0) & = & 4\, \frac{s^2_\phi c^2_\phi}{s_\beta
    c^2_\beta}\, \frac{\Im m b^2\Re e b^2}{v^2}\, \Big[\, I(M_1,M_1)\, +\,
    I(M_2,M_2)\, \Big]\, ,\\
  \label{hAc}
\Pi^{hA}_{(c)} (0) & = & \frac{c_\beta T_A}{v}\ =\ 
    2\, \frac{s_\phi c_\phi}{s_\beta}\, 
    \frac{\Im m b^2}{v^2}\, (M^2_1 - M^2_2)\, I(M_1,M_2)\, .
\end{eqnarray}
Adding the self-energy  expressions in Eqs.\ (\ref{hAa})--(\ref{hAc}),
we find
\begin{eqnarray}
  \label{PihA}
\Pi^{hA} (0) &=& 4\, \frac{s^2_\phi c^2_\phi}{s_\beta c^2_\beta}\,
    \frac{\Im m b^2\Re e b^2}{v^2}\, \Big[\, I(M_1,M_1)\, +\,
    I(M_2,M_2)\, -\, 2I(M_1,M_2)\, \Big]\nonumber\\
& = & \frac{s^2_\phi c^2_\phi}{s_\beta c^2_\beta}\,
    \frac{\Im m b^2\Re e b^2}{2\pi^2 v^2}\, \Big[\, 
    \frac{M^2_1 + M^2_2}{2(M^2_1 - M^2_2)}\, 
    \ln\Big(\frac{M^2_1}{M^2_2}\Big)\ -\ 1\, \Big]\, .
\end{eqnarray}
It is obvious  that the $hA$ self-energy becomes  UV finite only after
the CP-odd tadpole graph in Fig.\  4(c) has been included.  Similarly,
one may calculate  the CP-violating $HA$  transition and arrive  at an
analogous  UV-safe analytic expression.   Finally, we should emphasize
that  $\Pi^{hA}(0)$ shows   up a strong  non-decoupling  behaviour for
large mass   differences of the    charged scalars  $\chi^\pm_1$   and
$\chi^\pm_2$,    {\em i.e.},   $\Pi^{hA}(0)   \propto \Im  m f^2   \ln
(M_2/M_1)$   for $M_2/M_1\gg   1$   \cite{Aoki}.  The phenomenological
consequences  of the Abelian two-Higgs model  will be discussed in the
next section.

\setcounter{equation}{0}
\section{\label{sec:Concl}Discussion}

The existence of a sizeable  Higgs scalar-pseudoscalar mixing may have
profound implications for experiments of CP violation both at collider
and lower energies.  In  particular, a large CP-violating  $hA$ mixing
may  give rise to substantial  contributions to  the EDM's of neutron,
electron and muon. The experimental upper bounds on the EDM's of these
light fermions  are very   tight  \cite{PDG}: $d_{\rm n}   = 1.1\times
10^{-25}\ e\,$cm, $d_e  =   (-0.3\pm 0.8)\times 10^{-26}\   e\,$cm and
$d_\mu = (3.7\pm  3.4) \times 10^{-19}\  e\,$cm.\footnote[1]{There has
  been a recent proposal that next-round experiments at the Brookhaven
  National Laboratory may improve the accuracy of present measurements
  on  the muon  EDM  by six  orders of magnitude   \cite{BNL}.} On the
theoretical  side, Barr and Zee suggested  a mechanism \cite{BZ} which
is found to play a key role in enhancing the  EDM prediction in models
with CP   violation  in the  Higgs sector.   Even though  the Barr-Zee
mechanism occurs  at  two loops, it   is still very sensitive  to  the
mixing  of   scalar    and pseudoscalar   Higgs particles.     Shortly
afterwards,  Gunion and Wyler  extended this idea  by contemplating an
analogous quark chromo-EDM operator  \cite{Wyler}, which may  dominate
in the neutron EDM  over other CP-violating operators  as such of  the
CP-odd three-gluon moment introduced by Weinberg \cite{SW}.

Taking the afore-mentioned contributions into account, Hayashi {\em et
  al.}\ \cite{Hayashi}   have  constrained the  parameter   space of a
two-Higgs doublet model   with maximal  CP violation  \cite{SW}.  They
found that  the mass  splitting $\Delta   M$  between a scalar  and  a
pseudoscalar Higgs boson   due to a   tree-level $hA$ or $HA$   mixing
should not   be too  large.  Adapting   their results  to  the Abelian
two-Higgs model, we may estimate the upper limits
\begin{equation}
  \label{EDMlimit}
\frac{\Delta M}{M} \ <\ 0.10\,,\ 0.13\,,\ 0.24\, ,  
\end{equation}
for $M = (M_h+M_A)/2  = 200$, 400, and  600 GeV, respectively.  In the
above estimate, we have assumed that  $(M^2_h - M^2_A)/M^2 \ll 1$, and
$M_h,M_A \ll M_H$.   The proposed Brookhaven  experiment searching for
the muon  EDM \cite{BNL} might lower  further the upper bounds in Eq.\ 
(\ref{EDMlimit}), even up to one order of magnitude.

Since we   are interested in   confronting our theoretical predictions
with    the    phenomenological limits   on   a   $hA$  mixing in Eq.\
(\ref{EDMlimit}), it is more convenient to do so by reducing the large
number   of independent parameters  present  in  the Abelian two-Higgs
model.  For definiteness, we choose
\begin{equation}
  \label{data}
\tan\beta\ =\ \tan\phi\ =\ 1\, ,\qquad
\Re e\mu^2\ =\ -2\lambda_1 v^2_1\ =\ -\lambda_1 v^2\, .
\end{equation}
With the above choice of parameters, we have a degenerate $hA$ system,
namely  $M^2_A = M^2_h =  -2\lambda_1 v^2$, whereas the $H$-boson mass
is in general not fixed, {\em i.e.},  $M^2_H = -(\lambda_1 +\lambda_2)
v^2$.  The  mass   degeneracy of $h$   and $A$  is  then  lifted after
integrating out  the      charged scalar states   $\chi^\pm_1$     and
$\chi^\pm_2$.  The self-energy $\Pi^{hA}(0)$ in Eq.\ (\ref{PihA}) will
then quantify the loop-induced  mass difference $\Delta M$ between $h$
and $A$. To a good approximation, we obtain
\begin{equation}
  \label{DeltaM}
\frac{\Delta M}{M}\ \approx\ \frac{\Pi^{hA}(0)}{M^2}\ =\
(0.56,\ 2.05,\ 3.20,\ 4.10,\ 4.83)\times 10^{-2} \times \Big( \frac{\Im mb^2
\Re eb^2}{v^2 M^2}\Big)\, ,
\end{equation}
for ratios of charged scalar masses $M_2/M_1 = 2,\ 4,\  6,\ 8$ and 10,
respectively. Furthermore, in order to  retain the perturbative nature
of the  Higgs potential, the mass  parameter $|b|$  should be of order
$v$ and the  quartic couplings should not be  much larger  than unity. 
These two facts  allow us to deduce  the qualitative limit  $|\Im mb^2
\Re eb^2/(v^2 M^2)|  < 10$.  We then find  that  the parameter $\Delta
M/M$  measuring the radiative  mass splitting between  $h$ and $A$ may
adequately reach the  observable level of  10\%  for modest  values of
$|b|$  and charged-scalar-mass ratios, {\em e.g.},  for $|\Im mb^2 \Re
eb^2/(v^2  M^2)|  =  5$  and $M_2/M_1 =   4$.  Clearly,  more accurate
constraints on  the Abelian model may  be derived if a global analysis
of  all sensitive   low-energy observables,  such  as the  electroweak
oblique parameters $S$, $T$ and  $U$ \cite{PT}, is performed.  For the
more realistic case of a 2HDM, such  an analysis will crucially depend
on many other model details in the gauge sector.   We shall not embark
upon   this  topic here.   Instead,  it  may be   worth stressing that
scenarios with $M_h\approx  M_A$ may lead  to spectacular phenomena of
resonant CP  violation    through  scalar-pseudoscalar    mixing    at
high-energy $pp$, $e^+e^-$   and $\mu^+\mu^-$ colliders.  More details
may be found in \cite{APRL}.

It appears rather difficult  to determine experimentally the origin of
a  non-vanishing Higgs   scalar-pseudoscalar mixing  which generically
occurs   in  models with  extended  Higgs   sectors.    A CP-violating
Higgs-scalar  mixing could  arise either   spontaneously, due to   the
spontaneous  breakdown   of CP either  at the    tree level or through
quantum mechanical  effects, or explicitly  due to CP-violating Yukawa
interactions.  Here,  we have concentrated on  the latter alternative. 
Within  the framework of an  Abelian two-Higgs model, we have examined
the conditions  under which a  tree-level CP-invariant Higgs potential
can consistently  co-exist with other  CP-violating couplings, without
spoiling  renormalizability    at  high     orders.  Based   on    the
non-renormalization theorem, we have reached the conclusion that for a
tree-level Higgs potential, CP invariance  can be enforced by a global
U(1)$_Q$ symmetry  of the Peccei-Quinn type,  which can only be broken
softly, namely   by CP-violating operators  having dimensionality less
than four.  Within such  a  CP-violating scenario, the  CP-odd tadpole
renormalization induced by the tadpole graph of the pseudoscalar Higgs
boson  $A$  has  been  found   to  be  very important  to   render the
radiatively   generated $hA$  and  $HA$   transitions UV finite.   The
formalism developed in this   paper may apply   equally well  to  more
involved  theories,   such  as the   MSSM.    We defer the   study  of
scalar-pseudoscalar mixing in  the MSSM to a forthcoming communication
\cite{SUSYAP}.

\bigskip
\noindent {\bf Acknowledgments.}   I wish to thank Yannis  Semertzidis
for  a  useful information concerning  the  experimental status of the
muon EDM, as well as Gorazd Cvetic  and Ralf Hempfling for discussions
on issues of renormalization.

\newpage

\def\theequation{\Alph{section}.\arabic{equation}}
\begin{appendix}
\setcounter{equation}{0}
\section{\label{app:mass}Higgs-boson mass matrix}

Here, we will  discuss  the diagonalization  of the Higgs-boson   mass
matrix and the stability  conditions for the  Higgs potential in  Eq.\ 
(\ref{Lmu}).  In general, the Higgs-boson masses in this Abelian model
are obtained by diagonalizing the $4\times 4$ matrix
\begin{equation}
  \label{MHgen}
{\cal M}^2_H\ =\ \left(\begin{array}{cc} M^2_S & M^2_{SP} \\
                                     M^2_{PS} & M^2_P \end{array} \right)\, ,
\end{equation}
where $M^2_S$, $M^2_P$ and $M^2_{SP}$ are all $2\times2$ sub-matrices.
In  Eq.\  (\ref{MHgen}), the  general mass   matrix ${\cal M}^2_H$  is
written in  the weak basis spanned  by the  fields $H_1$, $H_2$, $A_1$
and $A_2$.    Thus,  the matrices  $M^2_S$ and   $M^2_P$  describe the
CP-conserving mass   transitions   $H_i\to  H_j$  and  $A_i\to   A_j$,
respectively,     whereas  $M^2_{SP}$   and   $M^2_{PS}$   contain the
CP-violating $H_i\to A_j$  and $A_i\to H_j$ transitions.  Since ${\cal
  M}^2_H$   is  symmetric, the matrices   $M^2_S$  and $M^2_P$ must be
symmetric as well, while  $M^2_{SP} = (M^2_{PS})^T$.   As we have seen
in Section \ref{sec:tad},  all entries of $M^2_{SP}$  are proportional
to the  CP-odd tadpole parameter $T_A$.  At  the tree level or  in the
CP-invariant limit   of the   theory,  we  have  $T_A  = 0$   and  the
diagonalization of $M^2_S$ and $M^2_P$ then proceeds independently.

To leading order, we adopt the limit of $T_A\to  0$ in the mass-matrix
diagonalization.  We start considering the mass matrix for the CP-even
Higgs scalars
\begin{equation}
  \label{M2S}
M^2_S\ =\  \left(\begin{array}{cc} 
-2\lambda_1 v^2_1 + \tan\beta\,\Re e \mu^2 - T_{H_1}/v_1 & 
                                    -\lambda_{34}v_1v_2 - \Re e \mu^2 \\
-\lambda_{34}v_1v_2 - \Re e \mu^2 &
-2\lambda_2 v^2_2 +\cot\beta\,\Re e\mu^2 - T_{H_2}/v_2\end{array} \right)\, , 
\end{equation}
where the tadpole  parameters $T_{H_1}$ and  $T_{H_2}$  are defined in
Eqs.\ (\ref{TH1}) and  (\ref{TH2}), respectively.  After diagonalizing
$M^2_S$ through the orthogonal transformation in Eq.\ (\ref{H1H2}), we
obtain the mass eigenstates $h$ and $H$. Their physical masses and the
respective Higgs-scalar mixing are related to the weak  parameters of 
the  Higgs potential as follows:
\begin{eqnarray}
  \label{MhH}
-2\lambda_1 v^2_1 + \tan\beta\,\Re e \mu^2 - T_{H_1}/v_1 & = &
M^2_h c^2_\theta\ +\ M^2_H s^2_\theta\, ,\nonumber\\
-2\lambda_2 v^2_2 + \cot\beta\,\Re e \mu^2 - T_{H_2}/v_2 &=&
M^2_h s^2_\theta\ +\ M^2_H c^2_\theta\, ,\nonumber\\
-\lambda_{34} v_1 v_2\, -\, \Re e\mu^2 &=& (M^2_H - M^2_h) 
                                                 s_\theta c_\theta\, .  
\end{eqnarray}
{}From Eq.\ (\ref{MhH}), we see  that stability of the Higgs potential
can naturally be achieved for negative values of the quartic couplings
$\lambda_1$ and  $\lambda_2$, whereas  $\lambda_{34}$ may  have either
sign.  From   an  analogous analysis  of  $M^2_P$,  we find  that  the
parameter $\Re e\mu^2$ must always  be  positive. More explicitly,  we
have for the mass matrix of pseudoscalars
\begin{equation}
  \label{M2P}
M^2_P\ =\  \left(\begin{array}{cc} 
 \tan\beta\,\Re e \mu^2 - T_{H_1}/v_1 & - \Re e \mu^2 \\
- \Re e \mu^2 & \cot\beta\,\Re e \mu^2 - T_{H_2}/v_2 \end{array} \right)\, . 
\end{equation}
The  mass  matrix  $M^2_P$ can   be   diagonalized via the  orthogonal
transformation  of  the  weak  fields $A_1$ and   $A_2$  given in  Eq.\
(\ref{G0A}). In the mass basis, $M^2_P$ reads:
\begin{equation}
  \label{M2Phat}
\widehat{M}^2_P\ =\  \left( \begin{array}{cc} 
-\, \frac{\displaystyle c_\beta T_{H_1} + s_\beta T_{H_2}}{\displaystyle  v} & 
\frac{\displaystyle s_\beta T_{H_1} - c_\beta T_{H_2}}{\displaystyle v}\\
\frac{\displaystyle s_\beta T_{H_1} - c_\beta T_{H_2}}{\displaystyle v}&
\quad\frac{\displaystyle \Re e \mu^2}{\displaystyle s_\beta c_\beta} - 
\frac{\displaystyle s_\beta\tan\beta\, T_{H_1} +
c_\beta\cot\beta\, T_{H_2}}{\displaystyle v} 
\end{array} \right)\, . 
\end{equation}
It is now easy to see that at the  tree level, $\widehat{M}^2_P$ has a
massless eigenvalue  corresponding to the  true Goldstone  boson $G^0$
and a massive one related to the CP-odd Higgs boson $A$, {\em i.e.},
\begin{equation}
  \label{MassA}
M^2_A\ =\ \frac{\Re e \mu^2}{s_\beta c_\beta}\ .
\end{equation}
Since  $M^2_A$ should  always be positive  for  a  stable theory,  the
latter implies that $\Re e \mu^2 > 0$.

\end{appendix}

\newpage

\centerline{\Large{\bf Figure captions }}
\vspace{-0.2cm}
\newcounter{fig}
\begin{list}{\rm {\bf Fig. \arabic{fig}: }}{\usecounter{fig}
\labelwidth1.6cm \leftmargin2.5cm \labelsep0.4cm \itemsep0ex plus0.2ex }

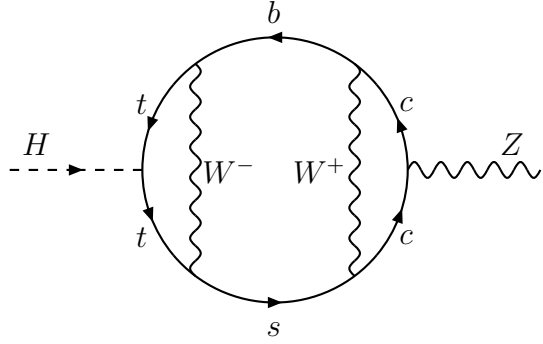
\item[{\bf  Fig.\ 1:}]  Typical   three-loop graph  giving rise  to  a
  non-vanishing $HZ$ mixing  in the SM.  The  remaining graphs  may be
  obtained by attaching $H$ and $Z$ in all possible  ways to the quark
  and $W$-boson lines.

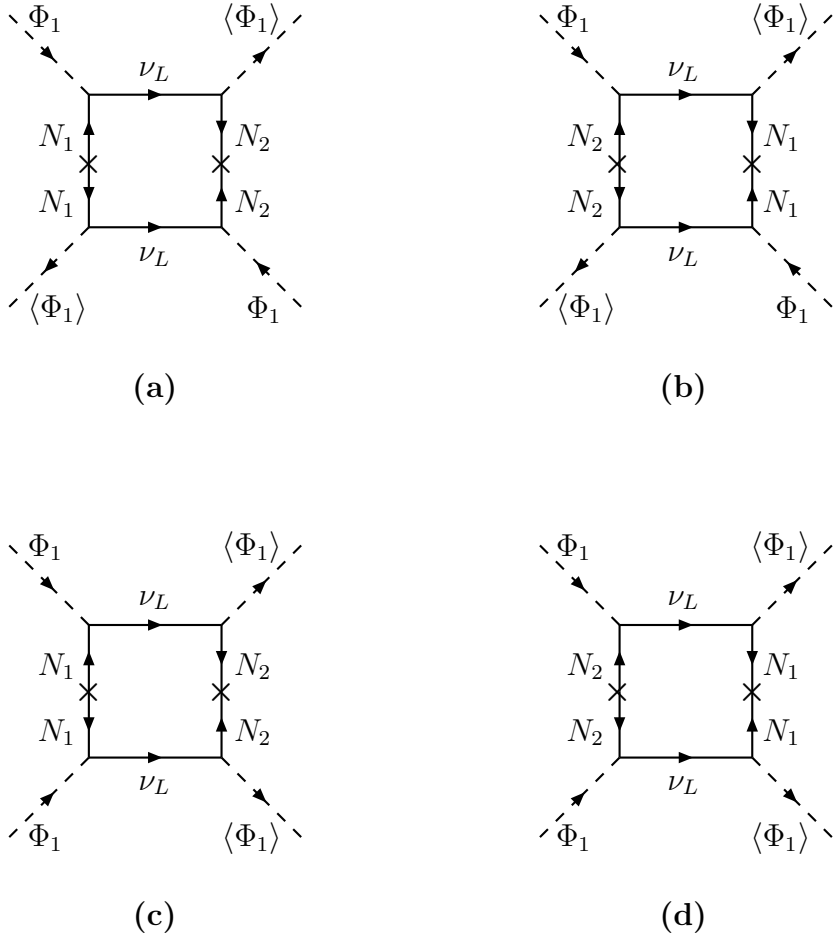
\item[{\bf Fig.\ 2:}] One-loop flavour graphs responsible for a
  non-vanishing $HA$ mixing in the Majorana-neutrino model.

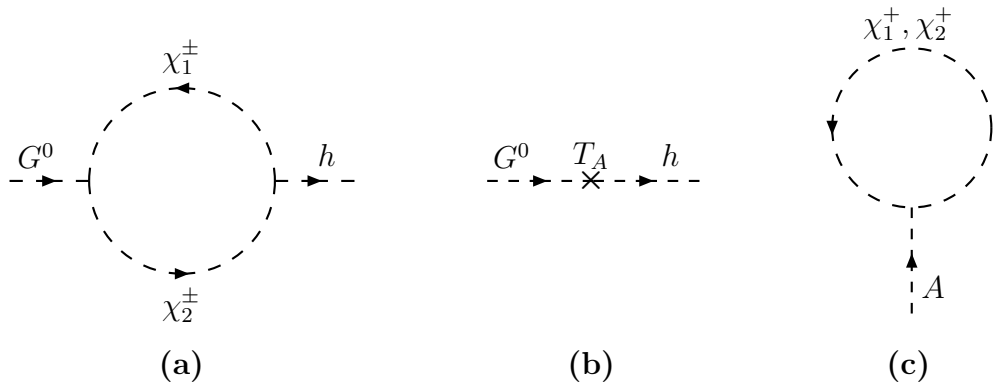
\item[{\bf Fig.\ 3:}] Diagrams pertinent to the $G^0h$ mixing: (a)
One-loop self-energy graph, (b) CP-odd tadpole renormalization,  
(c) Tadpole graph of the $A$ boson.

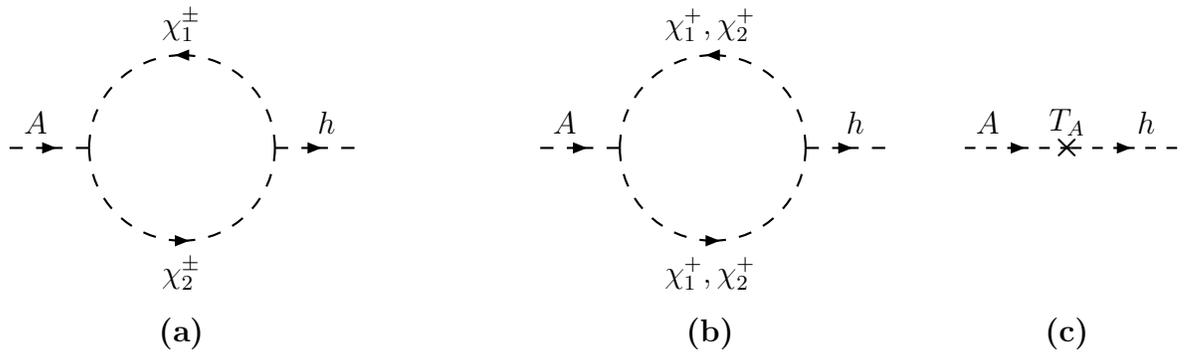
\item[{\bf Fig.\ 4:}] Diagrams pertinent to the $hA$ mixing.

\end{list}

\newpage

%******************************************************************
%%%Figure 1
%******************************************************************

\begin{center}
\begin{picture}(300,200)(0,0)
\SetWidth{0.8}

\DashArrowLine(0,100)(50,100){4}\Text(10,110)[]{$H$}
\Photon(150,100)(200,100){3}{5}\Text(190,110)[]{$Z$}
\ArrowArc(100,100)(50,0,40)\Text(150,125)[]{$c$}
\ArrowArc(100,100)(50,40,140)\Text(100,160)[]{$b$}
\ArrowArc(100,100)(50,140,180)\Text(50,125)[]{$t$}
\ArrowArc(100,100)(50,180,220)\Text(50,75)[]{$t$}
\ArrowArc(100,100)(50,220,320)\Text(100,40)[]{$s$}
\ArrowArc(100,100)(50,320,360)\Text(150,75)[]{$c$}
\Photon(70,140)(70,60){2}{7}\Text(73,100)[l]{$W^-$}
\Photon(130,140)(130,60){2}{7}\Text(127,100)[r]{$W^+$}

\Text(100,20)[t]{\bf Fig.\ 1}

\end{picture}
\end{center}

\vspace{0.5cm}
%******************************************************************
%%%Figure 2
%******************************************************************

\begin{center}
\begin{picture}(320,400)(0,0)
\SetWidth{0.8}

%%%

\DashArrowLine(0,380)(30,350){4}
\DashArrowLine(30,300)(0,270){4}
\DashArrowLine(80,350)(110,380){4}
\DashArrowLine(110,270)(80,300){4}
\ArrowLine(30,325)(30,300)\Text(30,325)[]{\boldmath $\times$}
\ArrowLine(30,325)(30,350)
\ArrowLine(30,350)(80,350)\ArrowLine(30,300)(80,300)
\ArrowLine(80,300)(80,325)\Text(80,325)[]{\boldmath $\times$}
\ArrowLine(80,350)(80,325)
\Text(25,310)[r]{$N_1$}\Text(25,335)[r]{$N_1$}
\Text(85,310)[l]{$N_2$}\Text(85,335)[l]{$N_2$}
\Text(55,360)[]{$\nu_L$}\Text(55,290)[]{$\nu_L$}
\Text(7,380)[l]{$\Phi_1$}\Text(7,270)[l]{$\langle\Phi_1\rangle$}
\Text(103,380)[r]{$\langle\Phi_1\rangle$}\Text(103,270)[r]{$\Phi_1$}

\Text(55,240)[]{\bf (a)}

%%%

\DashArrowLine(200,380)(230,350){4}
\DashArrowLine(230,300)(200,270){4}
\DashArrowLine(280,350)(310,380){4}
\DashArrowLine(310,270)(280,300){4}
\ArrowLine(230,325)(230,300)\Text(230,325)[]{\boldmath $\times$}
\ArrowLine(230,325)(230,350)
\ArrowLine(230,350)(280,350)\ArrowLine(230,300)(280,300)
\ArrowLine(280,300)(280,325)\Text(281,325)[]{\boldmath $\times$}
\ArrowLine(280,350)(280,325)
\Text(225,310)[r]{$N_2$}\Text(225,335)[r]{$N_2$}
\Text(285,310)[l]{$N_1$}\Text(285,335)[l]{$N_1$}
\Text(255,360)[]{$\nu_L$}\Text(255,290)[]{$\nu_L$}
\Text(207,380)[l]{$\Phi_1$}\Text(207,270)[l]{$\langle\Phi_1\rangle$}
\Text(303,380)[r]{$\langle\Phi_1\rangle$}\Text(303,270)[r]{$\Phi_1$}

\Text(255,240)[]{\bf (b)}

%%%

\DashArrowLine(0,180)(30,150){4}
\DashArrowLine(0,70)(30,100){4}
\DashArrowLine(80,150)(110,180){4}
\DashArrowLine(80,100)(110,70){4}
\ArrowLine(30,125)(30,100)\Text(30,125)[]{\boldmath $\times$}
\ArrowLine(30,125)(30,150)
\ArrowLine(30,150)(80,150)\ArrowLine(30,100)(80,100)
\ArrowLine(80,100)(80,125)\Text(80,125)[]{\boldmath $\times$}
\ArrowLine(80,150)(80,125)
\Text(25,110)[r]{$N_1$}\Text(25,135)[r]{$N_1$}
\Text(85,110)[l]{$N_2$}\Text(85,135)[l]{$N_2$}
\Text(55,160)[]{$\nu_L$}\Text(55,90)[]{$\nu_L$}
\Text(7,180)[l]{$\Phi_1$}\Text(7,70)[l]{$\Phi_1$}
\Text(103,180)[r]{$\langle\Phi_1\rangle$}
\Text(103,70)[r]{$\langle\Phi_1\rangle$}

\Text(55,40)[]{\bf (c)}

%%%

\DashArrowLine(200,180)(230,150){4}
\DashArrowLine(200,70)(230,100){4}
\DashArrowLine(280,150)(310,180){4}
\DashArrowLine(280,100)(310,70){4}
\ArrowLine(230,125)(230,100)\Text(230,125)[]{\boldmath $\times$}
\ArrowLine(230,125)(230,150)
\ArrowLine(230,150)(280,150)\ArrowLine(230,100)(280,100)
\ArrowLine(280,100)(280,125)\Text(281,125)[]{\boldmath $\times$}
\ArrowLine(280,150)(280,125)
\Text(225,110)[r]{$N_2$}\Text(225,135)[r]{$N_2$}
\Text(285,110)[l]{$N_1$}\Text(285,135)[l]{$N_1$}
\Text(255,160)[]{$\nu_L$}\Text(255,90)[]{$\nu_L$}
\Text(207,180)[l]{$\Phi_1$}\Text(207,70)[l]{$\Phi_1$}
\Text(303,180)[r]{$\langle\Phi_1\rangle$}
\Text(303,70)[r]{$\langle\Phi_1\rangle$}

\Text(255,40)[]{\bf (d)}

\Text(160,0)[t]{\bf Fig.\ 2}

\end{picture}
\end{center}

%******************************************************************
%%%Figure 3
%******************************************************************

\begin{center}
\begin{picture}(400,200)(0,0)
\SetWidth{0.8}

%%%

\DashArrowLine(0,100)(30,100){4}\Text(10,110)[]{$G^0$}
\DashArrowLine(100,100)(130,100){4}\Text(120,110)[]{$h$}
\DashArrowArc(65,100)(35,0,180){5}\Text(65,147)[]{$\chi^\pm_1$}
\DashArrowArc(65,100)(35,180,360){5}\Text(65,53)[]{$\chi^\pm_2$}

\Text(65,30)[]{\bf (a)}

%%%

\DashArrowLine(180,100)(220,100){4}\Text(190,110)[]{$G^0$}
\DashArrowLine(220,100)(260,100){4}\Text(250,110)[]{$h$}
\Text(220,100.5)[]{\boldmath $\times$}\Text(220,110)[]{$T_A$}

\Text(220,30)[]{\bf (b)}

%%%

\DashArrowArc(340,120)(30,0,360){5}\Text(340,160)[]{$\chi^+_1,\chi^+_2$}
\DashArrowLine(340,50)(340,90){4}\Text(345,60)[l]{$A$}

\Text(340,30)[]{\bf (c)}

\Text(200,0)[t]{\bf Fig.\ 3}

\end{picture}
\end{center}

\vspace{2.cm}
%******************************************************************
%%%Figure 4
%******************************************************************

\begin{center}
\begin{picture}(400,200)(0,0)
\SetWidth{0.8}

%%%

\DashArrowLine(0,100)(30,100){4}\Text(10,110)[]{$A$}
\DashArrowLine(100,100)(130,100){4}\Text(120,110)[]{$h$}
\DashArrowArc(65,100)(35,0,180){5}\Text(65,147)[]{$\chi^\pm_1$}
\DashArrowArc(65,100)(35,180,360){5}\Text(65,53)[]{$\chi^\pm_2$}

\Text(65,30)[]{\bf (a)}

%%%

\DashArrowLine(200,100)(230,100){4}\Text(210,110)[]{$A$}
\DashArrowLine(300,100)(330,100){4}\Text(320,110)[]{$h$}
\DashArrowArc(265,100)(35,0,180){5}\Text(265,147)[]{$\chi^+_1,\chi^+_2$}
\DashArrowArc(265,100)(35,180,360){5}\Text(265,53)[]{$\chi^+_1,\chi^+_2$}

\Text(265,30)[]{\bf (b)}

%%%

\DashArrowLine(360,100)(400,100){4}\Text(370,110)[]{$A$}
\DashArrowLine(400,100)(440,100){4}\Text(430,110)[]{$h$}
\Text(400,100.5)[]{\boldmath $\times$}\Text(400,110)[]{$T_A$}

\Text(400,30)[]{\bf (c)}

\Text(220,0)[t]{\bf Fig.\ 4}

\end{picture}
\end{center}

\end{document}